\documentclass[sigconf,screen,9pt]{acmart}
\usepackage{multirow}
\usepackage{graphicx}
\usepackage{ragged2e} 
\usepackage{flushend}
\usepackage{longtable} % 导入 longtable 包

\usepackage{fancyhdr}

\newcommand{\zdel}[1]{}

 \AtBeginDocument{%
   \providecommand\BibTeX{{%
     \normalfont B\kern-0.5em{\scshape i\kern-0.25em b}\kern-0.8em\TeX}}}

\copyrightyear{2024}
\acmYear{2024}
\setcopyright{acmlicensed}\acmConference[DAC '24]{61st ACM/IEEE Design Automation Conference}{June 23--27, 2024}{San Francisco, CA, USA}
\acmBooktitle{61st ACM/IEEE Design Automation Conference (DAC '24), June 23--27, 2024, San Francisco, CA, USA}
\acmDOI{10.1145/3649329.3656236}
\acmISBN{979-8-4007-0601-1/24/06}

\begin{document}

\title{DH-TRNG: A Dynamic Hybrid TRNG with Ultra-High Throughput and Area-Energy Efficiency}

\begin{abstract}
As a vital security primitive, the true random number generator (TRNG) is a mandatory component to build roots of trust for any encryption system. However, existing TRNGs suffer from bottlenecks of low throughput and high area-energy consumption. In this work, we propose DH-TRNG, a dynamic hybrid TRNG circuitry architecture with ultra-high throughput and area-energy efficiency. Our DH-TRNG exhibits portability to distinct process FPGAs and passes both NIST and AIS-31 tests without any post-processing. The experiments show it incurs only 8 slices with the highest throughput of 670Mbps and 620Mbps on Xilinx Virtex-6 and Artix-7, respectively. Compared to the state-of-the-art TRNGs, our proposed design has the highest $\mathrm{\frac{Throughput}{Slices\cdot Power} } $ with 2.63$\times$ increase.
\end{abstract}

%\begin{CCSXML}
%<ccs2012>
%<concept>
%<concept_id>10010583.10010600</concept_id>
%<concept_desc>Hardware~Integrated circuits</concept_desc>
%<concept_significance>500</concept_significance>
%</concept>
%<concept>
%<concept_id>10002978.10003001.10003599</concept_id>
%<concept_desc>Security and privacy~Hardware security implementation</concept_desc>
%<concept_significance>500</concept_significance>
%</concept>
%</ccs2012>
%\end{CCSXML}


\author{Yuan Zhang}
\affiliation{%
  \institution{Hunan University}
  \city{Changsha}
  \country{China}}
\email{zhangyuanzy@hnu.edu.cn}

\author{Kuncai Zhong}
\affiliation{%
  \institution{Hunan University}
  \city{Changsha}
  \country{China}}
\email{kczhong@hnu.edu.cn}

\author{Jiliang Zhang}
  % \thanks{*Corresponding author: Jiliang Zhang.}
   \authornote{Corresponding author: Jiliang Zhang.}
\affiliation{%
  \institution{Hunan University}
  \city{Changsha}
  \country{China}}
\email{zhangjiliang@hnu.edu.cn}


\maketitle

\section{Introduction}

Serving as a fundamental hardware security primitive, true random number generator (TRNG) is a crucial module in modern information security systems \cite{TIFS2018}. It extracts unpredictable inherent physical processes and generates completely unforeseeable random bitstreams to build roots of trust \cite{JSSC2022}, which makes it superior to the pseudo-random number generator driven by fixed algorithms in security.

%With the rapid development of information technology, system security has garnered increasing attention recently. As a hardware security primitive, the true random number generator (TRNG) establishes the most crucial trust root for every secure system\cite{TIFS2018}. It extracts inherent random physical processes to generate completely unpredictable random bitstreams. Compared to the pseudo-random number generator (PRNG) implemented by fixed algorithms, TRNG offers heightened security by its physical properties \cite{JSSC2022}. With the features of true randomness, unpredictability, and security, it has been widely applied in password generation, authentication protocol, key management, and random padding\cite{DAC2015}.
\begin{figure}[htbp]
 \vspace{10pt}
    \centering
    \includegraphics[scale=0.48]{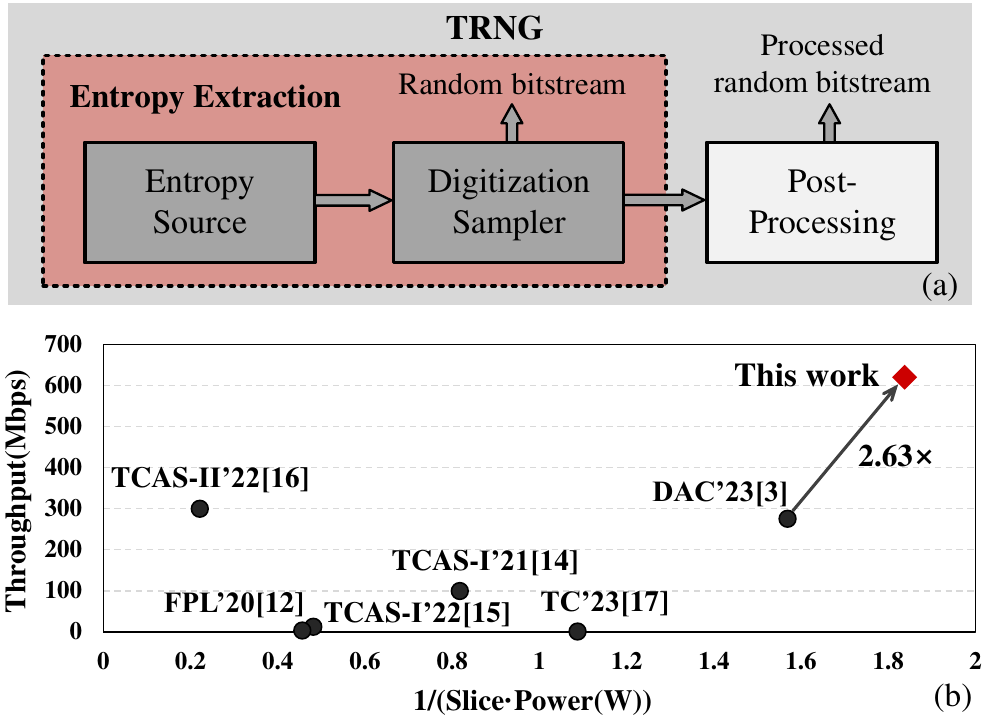}
    \vspace{-5 pt}
    \caption{(a) TRNG architecture. (b) Comparison with state-of-the-art TRNGs.}
    \vspace{-7pt}
    \label{Figure 1}
\end{figure}

Typically, a TRNG consists of entropy sources, digitization sampler, and post-processing \cite{DAC2023}, as shown in Figure~\ref{Figure 1}(a). The entropy source generates random events by extracting physical processes, the digitization sampler converts analog quantities into random sequences, and the post-processing transforms raw random numbers into bitstreams that satisfy randomness criteria. For these components, the quality of entropy extraction determines the necessity for post-processing and the performance of TRNG, thus occupying a central position in TRNG designs.

%In general, the usage of post-processing and the performance of TRNGs mainly depend on the quality of entropy source extraction. Hence, how to efficiently extract entropy sources is pivotal in TRNG design.

Due to the flexibility and application potential, the Field Programmable Gate Array (FPGA) has been the most prevalent hardware platform for implementing TRNGs. However, the current FPGA-compatible digital TRNGs still suffer from insufficient throughput and area-energy efficiency. 
The most crucial metric for optimizing TRNGs, denoted as $\mathrm{\frac{Throughput}{Slices\cdot Power}}$, presently only attains its maximum value of 432.97 in the recent work \cite{DAC2023}. 
%The $\mathrm{\frac{Throughput}{Slices\cdot Power}}$ is the core metric for optimizing TRNG, and the current highest value of $\mathrm{\frac{Throughput}{Slices\cdot Power}}$ is 432.97 in work \cite{DAC2023}. 
It is inadequate to meet the growing demand for substantial amounts of encrypted data and a high rate of random number generation in emerging scenarios, such as confidential computing, trusted execution environments, blockchain digital signatures, and edge computing.

To address the above issues, we propose DH-TRNG, a dynamic hybrid true random number generator. To effectively improve the throughput, it exploits multiple entropy sources by dynamic switching of loop state to support higher sampling frequencies. Moreover, it utilizes two reinforcement strategies to further enhance output randomness and reduce entropy unit usage while minimizing the incurred area.

Our main contributions are summarized as follows.

\begin{itemize}

\item We propose a hybrid entropy unit capable of dynamically quantifying the randomness of both jitter and metastability. We verify the randomness improvement through theoretical derivation and experiments.

\item We propose coupling and feedback strategies to further improve output randomness and reduce entropy unit usage. Then, we introduce a multistage sampling array that generates bitstreams capable of passing the statistical tests without any post-processing. 

\item We optimize the circuitry area to make the proposed TRNG minimized and portable for FPGAs with various processes. Prominently, it exhibits ultra-high $\mathrm{\frac{Throughput}{Slices\cdot Power}} $ with promising application prospects. The source code is available at \url{https://github.com/zy-cshnu/DH-TRNG}.

% {https://github.com/blind-for-review}
\end{itemize}

The experimental results demonstrate that the proposed TRNG architecture consumes a total of 8 slices, comprising 23 LUTs, 4 MUXs, and 14 DFFs. In Virtex-6 and Artix-7, our proposed design exhibits an ultra-high throughput of 670Mbps and 620Mbps with a low power consumption of 0.126W and 0.068W, respectively. As depicted in Figure~\ref{Figure 1}(b), compared to the state-of-the-art TRNGs, our DH-TRNG has the highest $\mathrm{\frac{Throughput}{Slices\cdot Power}}$ with 2.63$\times$ prominent increase. 

% The paper is organized as follows. In Section II, we provide the background and related work. The concrete circuit and principles of HE-TRNG are illustrated in Section III. We present the experimental results and comparison in Section IV. Finally, we conclude the paper in Section V.

\section{Background and Related Work}
\label{Section: Background and related work}

This section introduces the randomness principle of jitters and metastability, and presents the related work.

\subsection{ Oscillation Jitter }
\label{Section: Oscillation Jitter}
\begin{figure}[htbp]
% \vspace{-8pt}
    \centering
    \includegraphics[scale=0.52]{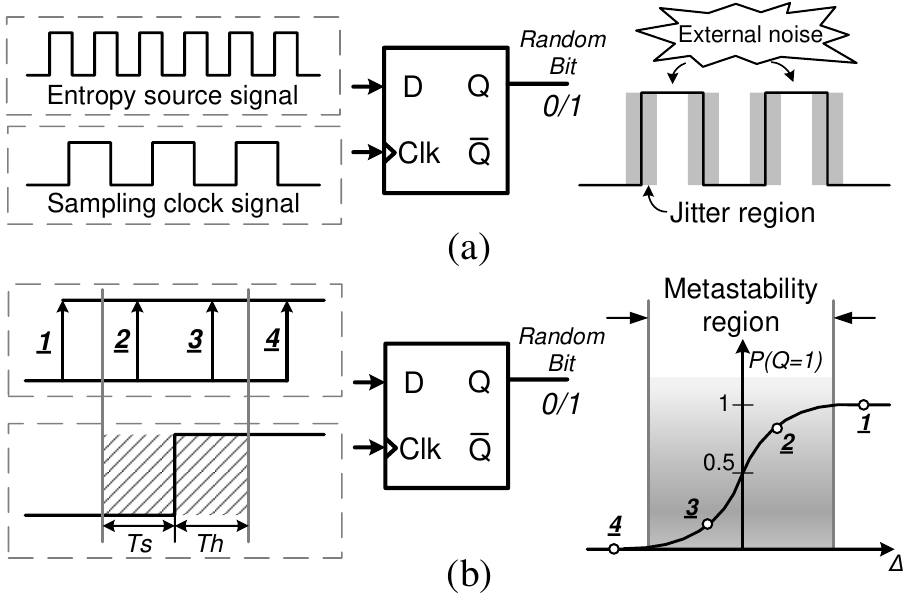}
    % \vspace{-7pt}
    \caption{(a) Randomness extraction of jitters. (b) Randomness extraction of metastability.}
    \label{Figure 2}
     \vspace{-10pt}
\end{figure}

Jitter is a typical entropy source induced by external random physical processes, such as thermal noise, scattering noise, power supply fluctuation, and environmental variation, in ring oscillators (ROs).
It manifests as uncertain phase noise in the frequency domain. Typically, in the jitter extraction process shown in Figure~\ref{Figure 2}(a), a low-frequency clock signal samples the high-frequency oscillation of entropy sources within a flip-flop to generate random bits. The relation between the phase noise and the oscillation ring order is expressed as \cite{Jitter}:
%Jitter is a typical entropy source that manifests as uncertain phase noise in the frequency domain, as shown in Figure 2(a). It is mainly induced by external random physical processes, such as thermal noise, scattering noise, power supply fluctuation, and environmental variation, in oscillation signals. Typically, in the process of jitter extraction, a low-frequency clock signal samples the high-frequency oscillation of entropy sources within a flip-flop to generate random bits. The relation between the phase noise and the oscillation ring order is expressed as \cite{Jitter}:
\begin{equation}
L_{min}\left \{ \Delta f \right \} =\frac{8N}{3\eta }\cdot \frac{KT}{P}\cdot \left ( \frac{V_{DD} }{V} +\frac{V_{DD} }{IR}  \right )\cdot \left ( \frac{f_{0}}{\Delta f}  \right )  ^{2},
\end{equation}
where $K$, $T$, $\eta $, $V_{DD}$, $V$, $I$, and $R$ are constants, $f_{0}$ is the frequency of the ring, $\Delta f$ denotes the offset frequency, $P$ is the power consumption, and $N$ is the order of the ring. 

Apparently, increasing the order $N$ can amplify phase noise $L_{min}$. Nevertheless, it will also reduce frequency $f_{0}$ and throughput \cite{Cui}. Thus, many related works focus on the trade-off schemes to improve performance. For instance, Cui et al. \cite{Cui} provided a multi-stage feedback RO that increased both $N$ and $f_{0}$. Lu et al. \cite{DAC2023} designed a multiphase sampler TRNG architecture, which increases the throughput with low resource overhead. However, their basic entropy units exhibit inherent randomness insufficiency, which limits the efficiency improvement of generating random sequences. 
%However, its basic entropy unit has inherent randomness deficiencies, which limits the improvement of the efficiency of generating random sequences.

 \subsection{Sampling Metastability}
In addition to jitter, metastability, which refers to the unpredictable random output of flip-flops, serves as another ideal entropy source for true random number generation \cite{JSSC2008, CHES2011}. As shown in scenarios 2 and 3 of Figure~\ref{Figure 2}(b), when the flip-flop samples unstable intermediate signals within the set/hold time,  unpredictable random bits are generated due to timing violations associated with metastability. 

Majzoobi et al. \cite{CHES2011} proved the probability of output settling onto ‘1’ can be accurately modeled by the Gaussian cumulative distribution function and central limit theorem, as expressed in Equation~\eqref{equation 2}: 
\begin{equation}
P(out=1)=Q\left ( \frac{\Delta  }{\sigma }  \right ) ,\\
Q(x)=\frac{1}{\sqrt{2\pi } }\int_{x}^{\infty } e^{\frac{-u^{2} }{2} } du ,
\label{equation 2}
\end{equation}
where $\Delta$ denotes the time difference between the 
sampling edge and the moment transition occurs, and $\sigma$ is proportional to the width of the setup/hold time window. 

Based on Equation~\eqref{equation 2}, reducing $\Delta$ can force the flip-flop into metastability by ensuring that sampling points fall between points 2 and 3 as much as possible, as depicted in Figure~\ref{Figure 2}(b). Inspired by this, Peng et al. \cite{TODAES2023} proposed RO-driven shift registers that align the delay of symmetric inputs to produce unpredictable equal probability outputs. Sala et al. \cite{TCAS1_2022} proposed latched-XOR cells configured by multiple excitation signals to achieve randomness improvement with low hardware area. Nevertheless, both their throughput is constrained due to limitations in entropy collection rate.

% \vspace{-4pt}
\section{The Proposed Dynamic Hybrid TRNG}

This section presents our proposed design. First, we introduce the proposed dynamic hybrid entropy unit. Then, we present two reinforcement strategies for enhancing randomness. Finally, we provide the overall circuitry architecture and the area optimization.

%In this section, we first propose and analyze the hybrid-entropy unit structure. Then, we introduce two reinforcement strategies for enhancing randomness. Finally, overview circuitry and area optimization are presented.

% \vspace{-4pt}
\subsection{Dynamic Hybrid Entropy Unit}

\begin{figure}[htbp]
    \centering
    \includegraphics[scale=0.58]{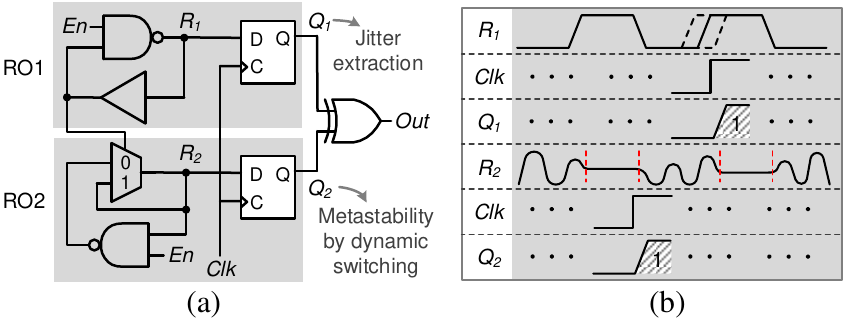}
     \vspace{-5 pt}
    \caption{(a) Dynamic hybrid unit structure. (b) Randomness extraction principle.}
    \label{Figure 3}
% \vspace{-12 pt}
\end{figure}

As discussed in Section~\ref{Section: Oscillation Jitter}, the randomness of ROs is affected by a trade-off between their order and oscillation frequency. 
To explore the optimal solution, we experimentally analyze the randomness of ROs with different orders. We utilize NIST SP800-90B to test the parallel XORed ROs with a 100MHz sampling frequency. As shown in Table~\ref{Table 1}, 9-stage ROs yield the highest min-entropy.
%To explore the optimal solution, we experimentally analyze the randomness of ROs with different orders. The results in Table 1, obtained through parallel XOR with a 100MHz sampling frequency, reveal that 9-stage ROs yield the highest min-entropy.
However, their long transmission delays make them cannot meet the throughput improvement target. In order to satisfy both high-frequency properties and sufficient randomness, we propose a dynamic hybrid entropy unit as shown in Figure~\ref{Figure 3}(a).

%There is still a lack of entropy source structure with both high-frequency property and adequate randomness. To address this issue, we propose a dynamic hybrid entropy unit as shown in Figure 3(a).

%The randomness of ROs is highly correlated with their order and oscillation frequency, which cannot be simultaneously increased, as expressed in section II. To find the optimal solution, we experimentally analyze the randomness across different orders of ROs in Fig. 3. In each set of experiments, a 100 MHz clock samples all types of ROs, and the outputs are 15 parallels XORed. The results shown in Fig.3 indicate that 9-order ROs provide the highest min-entropy. Nevertheless, due to their long transmission delay, the oscillation frequency cannot meet the aim of throughput improvement. Therefore, we employ low-order rings to ensure high oscillation frequency. In our design, to compensate for the effect of jitter accumulation, we propose a hybrid entropy unit as an entropy source with significantly improved randomness. 

%Then, we. Finally, we optimize the overall structure area of our HE-TRNG to minimize resource overhead. 

%图3和图5合并，或者使用表格代替

\begin{table}[htbp]
\centering
\renewcommand{\arraystretch}{1} % 设置行距为1.5倍，默认为1.0
 \vspace{-4pt}
\caption{Randomness test of different order oscillation rings.}
\label{Table 1}
\tabcolsep=2pt
\vspace{-4 pt}
\scalebox{0.95}
{ 
\begin{tabular}{ccccccc}
\hline
Stage number & 2      & 3      & 4      & 5      & 6      & 7                          \\\hline
Min-entropy  & 0.9737 & 0.9733 & 0.9756 & 0.9776 & 0.9783 & 0.9831                     \\ \hline \hline
Stage number & 8      & 9      & 10     & 11     & 12     & \multicolumn{1}{c}{13}     \\ \hline 
Min-entropy  & 0.9860  & 0.9871 & 0.9842 & 0.9837 & 0.9788 & \multicolumn{1}{c}{0.9735} \\ \hline
\end{tabular}}
% \vspace{-7 pt}
\end{table}

% \begin{figure}[htbp]
%     \centering
%     \includegraphics[scale=0.42]{Figure/fig5_ro_test.pdf}
%     \vspace{-8pt}
%     \caption{Randomness test of different order oscillation rings}   
% \end{figure}

%The randomness of multi-order XOR sampling is related to the order and the frequency of rings, as expressed in equation (1). we test the randomness of oscillation rings with variant orders using a 100 MHz clock, as shown in Figure 3. The results indicate that 9-order oscillation rings achieve the highest min-entropy. Nevertheless, the improvement of throughput requires lower-order rings. To achieve the effects of both high frequency and randomness, we propose a hybrid entropy unit, which extracts jitters while simultaneously quantifying the randomness introduced by metastability. The entropy value is effectively strengthened. 

For the proposed entropy unit, two independent oscillation rings RO1 and RO2 are enabled by signal $En$ to generate high-frequency oscillation signals. Two D flip-flops sample nodes $R_{1}$ and $R_{2}$, producing the final random output through XOR. Due to the influence of external noise, uncertain jitter regions are generated in the transition edge of $R_{1}$, as shown in Figure~\ref{Figure 3}(b). When the clock's positive edge samples this offset signal, the randomness of jitters is quantified, producing an unpredictable output $Q_{1}$. 

Meanwhile, $R_{1}$ serves as the activation signal for the MUX in RO2, causing RO2 to realize random dynamic switching between a holding loop and an inverter loop. When $R_{1}$=0, $R_{2}$ is in the oscillation region. A high-frequency oscillation signal smoothens the original square wave signal and generates short pulses, which expand the transition moment in the time domain. When $R_{1}$=1, $R_{2}$ is in the holding region, where a rapid transition signal is randomly locked at an uncertain subthreshold state. Thus, the signal collection of $Q_{2}$ greatly amplifies the probability of metastability, and the final dynamic hybrid $Out$ exhibits considerable randomness.
%Moreover, this dynamic path switching realizes random state transitions. 

% \vspace{-6 pt}

Next, we present the formal verification of the randomness increase. Ideal random numbers imply an equal generation probability of `0' and `1'. When sampling occurs within the holding region, the random subthreshold signal $R_{2}$ causes $\Delta$ in Equation~\eqref{equation 2} to be 0. Hence, the probability of $Q_{2}$ nears $\frac{1}{2}$. According to \cite{XOR2008}, the expected value of $Out$ is expressed as:
% \vspace{-6 pt}
\begin{equation}
E_{Out}=E(Q_{1}\oplus Q_{2})=\frac{1}{2} -2(\mu _{1} -\frac{1}{2})(\mu _{2} -\frac{1}{2}),
\end{equation}
where $\mu_{1}$ and $\mu_{2}$ are the expected values of $Q_{1}$ and $Q_{2}$, respectively. In the holding region, $R_{1}$ is the definite value 1, and $\mu_{2}\approx \frac{1}{2}$. Thus, the expected value is close to $\frac{1}{2}$, exhibiting near-ideal randomness.

At the same time, when sampling takes place within  the oscillation region, the $n$-order XOR expected value of $Out$ is calculated by Equation~\eqref{equation 4}:
% \vspace{-3 pt}
\begin{equation}
E_{Out_n}=E(Q_{1} \oplus Q_{2} )_n= \frac{1}{2} \left ( 1+\left ( \left ( 1-2\mu_{1}   \right ) \left ( 1-2\mu _{2}  \right )  \right )^{\frac{n}{2} }   \right ).
\label{equation 4}
\end{equation}
Given that $\mu _{1},\mu _{2}\in(0,1)$ implies $ 1-2\mu_{1},1-2\mu_{2}\in(-1,1)$, the expected value will rapidly converge to $\frac{1}{2}$ with an increase in the XOR number. Thus, employing more parallel entropy units can eliminate the bias and improve the randomness.

Finally, we show the randomness coverage of the proposed unit, which is a vital metric for evaluating the
entropy source. According to \cite{TCAS1_2017}, the randomness coverage of our proposed dynamic hybrid entropy unit through multi-level XOR is derived as: 
% \vspace{-5 pt}
\begin{equation}
P_{rand} =1-\prod_{i=1}^{n} \left (  (1-\frac{2aw_{i} }{T_{ro_{i} } } )(1-(\tau +2\varepsilon f_{i}   ))\right ),
\label{equation 5}
\end{equation}
where $n$ is the number of XOR. $a$ and $w_{i}$ are the probability and width of jitters in RO1. $T_{ro_{i}}$ denotes its oscillation period. $\tau$ is the probability of sampling subthreshold signals at the holding region in RO2. $\varepsilon$ and $f_{i}$ denote the transition edge width and oscillation frequency at the oscillation region. 

As shown in Equation~\eqref{equation 5}, the values of cumulative multiplication are greatly minimized, and the random coverage by multiple XORs is nearly closer to 1, indicating a significant randomness increase of the proposed design. 
Based on it, we also do some experiments to show the strength. As depicted in Table~\ref{Table 2}, compared to 9-stage ROs, dynamic hybrid entropy units exhibit higher min-entropy. Thus, our proposed entropy units successfully achieve both increased randomness and high-frequency characteristics.

%We tested and compared the randomness of dynamic hybrid entropy units with 9-stage ROs. The results in Table 2 demonstrate that our proposed entropy source structure exhibits higher min-entropy.
%用表个来代替
% \vspace{-7 pt}
\begin{table}[htbp]
\centering
\renewcommand{\arraystretch}{1} % 设置行距为1.5倍，默认为1.0
\vspace{-4 pt}
\caption{Comparison of entropy units with 9-stage ROs.}
\label{Table 2}
\vspace{-6 pt}
\scalebox{0.95}
{ 
\begin{tabular}{cccccc}
\hline
XOR number & 9      & 10     & 11     & 12     & 13     \\ \hline
Entropy units      & 0.9765 & 0.9803 & 0.9830  & 0.9836 & 0.9853 \\
9-stage ROs & 0.9705 & 0.9751 & 0.9779 & 0.9801 & 0.9813 \\ \hline
\hline
XOR number & 14     & 15     & 16     & 17     & 18     \\\hline
Entropy units      & 0.9868 & 0.9885 & 0.9896 & 0.9903 & 0.9912 \\
9-stage ROs & 0.9849 & 0.9871 & 0.9873 & 0.9886 & 0.9891 \\ \hline
\end{tabular}}
 \vspace{-5 pt}
\end{table}

% \begin{figure}[htbp]
%     \centering
%     \includegraphics[scale=0.45]{Figure/fig6.pdf}
%     \vspace{-8pt}
%     \caption{Randomness comparison of hybrid entropy units and oscillation rings.}   
% \end{figure}
% \vspace{-3 pt}
\subsection{Coupling and Feedback Strategies} 
% \vspace{-2 pt}
Increasing the employed entropy units enhances randomness but leads to higher hardware overhead and power consumption. To improve output randomness while reducing the entropy unit usage, we further apply two reinforcement strategies, which are coupling and feedback strategies.

 \vspace{-8 pt}
\begin{figure}[htbp]
    \centering
    \includegraphics[scale=0.47]{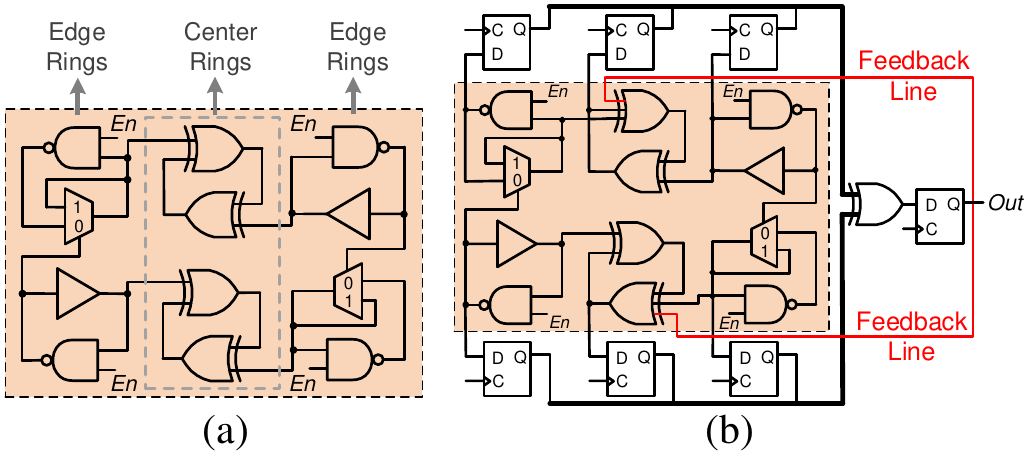}
     \vspace{-5pt}
    \caption{(a) Coupling strategy. (b) Feedback strategy.} 
    \label{fig 4}
 \vspace{-7 pt}
\end{figure}
In the coupling strategy, two dynamic hybrid entropy units are reversely inserted into two 2-stage XOR rings and create a nested coupling structure, as shown in Figure~\ref{fig 4}(a). It involves two central rings and four edge rings. Due to the characteristics of XOR gates, the logic of central rings is variable with inputs. Thus, signals within the central rings are activated by the oscillation signals with random phase noise on both sides. 
Furthermore, in the central rings, there are independent jitters and overlaps of different frequencies generated by the edge rings in the time domain. 
This leads to the logical mode of the central ring undergoing disorderly switching, and further causes the gate-level signals within it to exhibit nonperiodic random flips during propagation. 

%Hence, the oscillation signal of the central ring displays significant randomness. 

In the feedback strategy, the final output is fed back into central rings to randomize the initial state, as shown in Figure~\ref{fig 4}(b). For this strategy, multiple D flip-flops sample independent signals generated by various oscillation rings, the sampled random signals are collected by an XOR gate to produce the final output, and an additional flip-flop is introduced to transmit $Out$ into the central rings through feedback lines. This will cause random phases to switch in the oscillation signals. Thus, the high-entropy $Out$ serves as an activation signal to renewedly randomize the initial state, and further boosts the final output randomness.

The polynomial of a feedback XOR ring is expressed as 
$f(x)={\textstyle \sum_{i=1}^{n}}x_{i}^{k(i)} $, where $n$ and $k(i)$ is the stage number of XOR gates and edge rings, respectively. The feedback polynomial of the central ring for achieving multi-ring convergence is $f(x)=x_{1}^{1} +x_{2}^{2} +x_{r}^{'}$,
where $x_{1}^{1}$ and $x_{2}^{2}$ denote signals of edge rings, and $x_{r}^{'}$ is the feedback of the final random output. Thus, coupling and feedback strategies implement the stacking of random signals and a completely random initial state, greatly improving the chaos of the overall system and enhancing the randomness quality of the final output. 
 \vspace{-4 pt}
\subsection{Overall Architecture and Area Optimization}
\begin{figure}[htbp]
    \centering
    \includegraphics[scale=0.57]{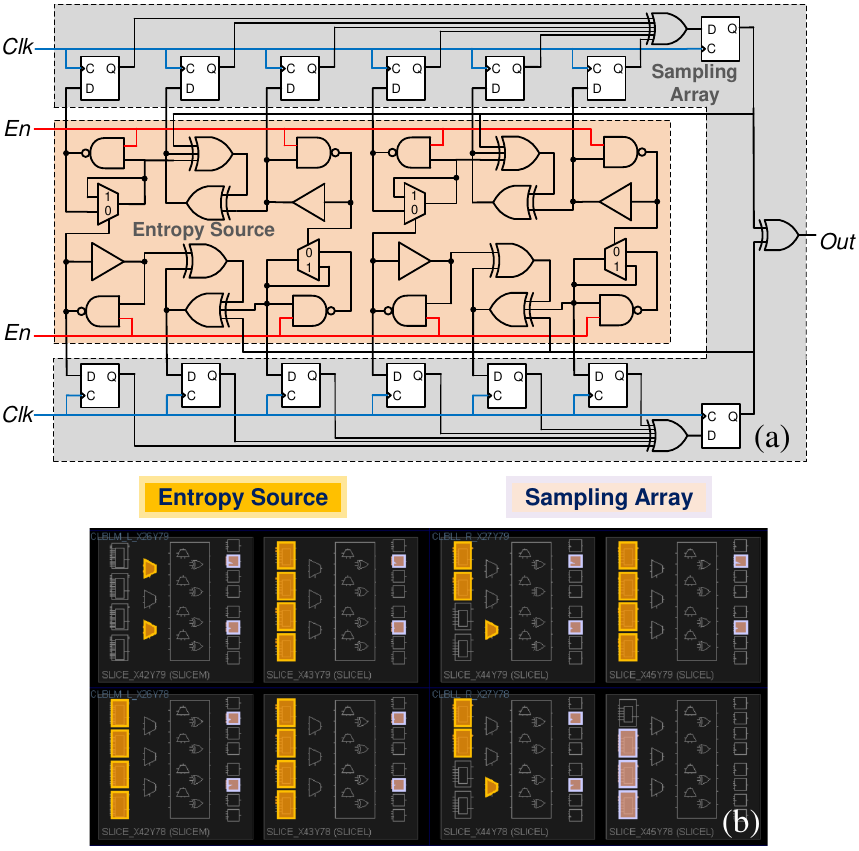}
     % \vspace{-4 pt}
    \caption{(a) Overall circuitry architecture of DH-TRNG. (b) Implementation with automated placement and routing.}
    \label{fig 5}
  % \vspace{-5 pt}
\end{figure}

Based on the aforementioned entropy units and strategies, we propose the DH-TRNG as shown in Figure~\ref{fig 5}(a).
%Figure 5(a) shows the overall architecture of the proposed DH-TRNG. 
In the entropy source, the proposed dynamic hybrid entropy units inserted into central XOR rings form two sets of identical nested coupling structures. In the multistage sampling array, all independent signals generated by the 12 rings are sampled separately through D flip-flops, ultimately generating random output via an XOR tree. Meanwhile, the final output is fed back to the input of the XOR rings in the entropy source to randomize the initial state. The entire circuitry is enabled by the signal $En$, and the $Clk$ is provided through the built-in Phase Locked Loop (PLL) in the FPGA. One bit true random number is produced at each clock cycle. Moreover, consider that our DH-TRNG can provide sufficient randomness support as discussed above. It allows for automatic layout and routing in FPGA without any manual intervention and performance loss. Hence, the proposed design also has high portability. 

%Due to sufficient randomness support, our DH-TRNG allows for automatic layout and routing in FPGA without any manual intervention and performance loss.

To reduce the area, we further optimize the
implementation of DH-TRNG. As the basic programmable unit of FPGA, one slice in Xilinx 6 serials or 7 serials FPGA contains four six-input LUTs, three MUXs, eight DFFs, and other arithmetic logic. All gate circuits can be implemented by configuring the LUT to corresponding logical functions. We constrain all gate cells by type to an appropriate position in a compact square slice array, as shown in Figure~\ref{fig 5}(b). Given the coordinates of the origin slice, only 8 slices are consumed, including 20 LUTs and 4 MUXs for entropy source structure, 14 DFFs and 3 LUTs for the sampling array. Moreover, the placement of the same type of gates and each slice can be flexibly adjusted. Thus, the area of our proposed DH-TRNG is greatly optimized.
% \vspace{-5 pt}

 \vspace{-2 pt}
\section{Experiment and Results}
\label{section: Performance Evaluation}
\begin{figure}[htbp]
    \centering
    \includegraphics[scale=0.46]{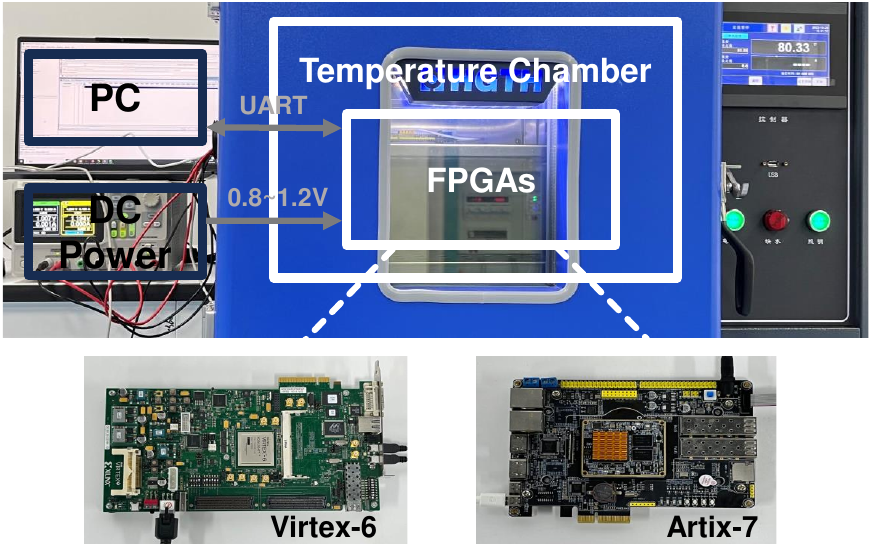}
     \vspace{-5 pt}
    \caption{Experimental platform.}
    \label{fig 6}
     \vspace{-10 pt}
\end{figure}

We implement the proposed DH-TRNG on two different process FPGAs to verify its portability, which are Xilinx Virtex-6 FPGA (xc6vlx240t with 45 nm process) and Xilinx Artix-7 FPGA (xc7a100t with 28 nm process). To evaluate its resistance to temperature and voltage variations, we set up an experimental platform illustrated in Figure~\ref{fig 6}, consisting of a temperature chamber, a DC power supply, a computer, and two FPGA boards. The performance of DH-TRNG is measured by a series of tests on the collected data, such as NIST test, AIS-31 test, deviation test, autocorrelation test, restart test, and PVT test. Finally, we conduct a comprehensive comparison of our proposed design with the state-of-the-art FPGA-based TRNG architectures.
% \vspace{-12 pt}
     \vspace{-10pt}
\subsection{Randomness Test}
    % \vspace{-2pt}
We perform three kinds of randomness tests on DH-TRNG.
    % \vspace{-5pt}
\subsubsection{NIST SP 800-22 Test}
It is a widely used test suite to verify the randomness of the random sequence, where the \textit{P-value} exceeding 0.01 indicates the sequences are approximately uniformly distributed. The \textit{Prop.} denotes the pass probability of the test item. We collect 30 sets of 1Mbit random numbers for each FPGA in a standard environment. As shown in Table~\ref{Table 3}, the test results illustrate that our proposed DH-TRNG passes all test items and exhibits good randomness. %\cite{NIST22}
 % \vspace{-5 pt}
\begin{table}[htbp]
\centering
\renewcommand{\arraystretch}{1} % 设置行距为1.5倍，默认为1.0
\vspace{-3 pt}
\caption{Results of NIST SP 800-22 Test.}
\label{Table 3}
\vspace{-4 pt}
\tabcolsep=2.7pt
\scalebox{0.9}
{ 
\begin{tabular}{ccccc}
\hline
\multirow{2}{*}{NIST SP 800-22} & \multicolumn{2}{c}{\textbf{Virtex-6}} & \multicolumn{2}{c}{\textbf{Artix-7}} \\ \cline{2-5} 
                                & \textit{P-value}             & \textit{Prop.}           & \textit{P-value}            & \textit{Prop.}           \\ \hline
Frequency                       & 0.739918            & 30/30           & 0.739918           & 30/30           \\
BlockFrequency                  & 0.100508            & 29/30           & 0.407091           & 29/30           \\
CumulativeSums*                 & 0.180952            & 30/30           & 0.462665           & 30/30           \\
Runs                            & 0.468595            & 30/30           & 0.178278           & 29/30           \\
LongestRun                      & 0.122325            & 30/30           & 0.213309           & 29/30           \\
Rank                            & 0.350485            & 30/30           & 0.350485           & 30/30           \\
FFT                             & 0.739918            & 30/30           & 0.468595           & 30/30           \\
NonOverlappingTemplate*         & 0.472949            & 30/30           & 0.477819           & 30/30           \\
OverlappingTemplate             & 0.671779            & 30/30           & 0.534146           & 30/30           \\
Universal                       & 0.350485            & 30/30           & 0.299251           & 29/30           \\
ApproximateEntropy              & 0.602458            & 30/30           & 0.804337           & 30/30           \\
RandomExcursions*               & 0.090867            & 17/17           & 0.029136           & 17/17           \\
RandomExcursionsVariant*        & 0.084577            & 17/17           & 0.043234           & 17/17           \\
Serial*                         & 0.390368            & 30/30           & 0.844760           & 30/30           \\
LinearComplexity                & 0.178278            & 29/30           & 0.407091           & 30/30           \\ \hline
\multicolumn{5}{l}{* The p-value is the average of the p-values of all subtests.}
\end{tabular}
}
\vspace{-7 pt}
\end{table}
% \vspace{-7 pt}

\subsubsection{NIST SP 800-90B Test}
This test focuses on testing the null hypotheses of a digital sequence and employs a distinct method for estimating the statistical distribution and min-entropy of the random sequences. We collect 30 sets of 1Mbit random numbers for each FPGA in a standard environment. The NIST SP 800-90B non-IID test results are shown in Table~\ref{Table 4}, and the min-entropy of the IID test is 0.994698 (Virtex-6) and 0.995966 (Atrix-7), respectively, demonstrating superior randomness.
%\cite{NIST90B}
\begin{table}[htbp]
\centering
\renewcommand{\arraystretch}{1} % 设置行距为1.5倍，默认为1.0
% \vspace{-2 pt}
\caption{Results of NIST SP 800-90B Test.}
\label{Table 4}
% \vspace{-7 pt}
\tabcolsep=2.7pt
\scalebox{0.9}
{
\begin{tabular}{ccccc}
\hline
\multirow{2}{*}{NIST SP 800-90B} & \multicolumn{2}{c}{\textbf{Virtex-6}} & \multicolumn{2}{c}{\textbf{Aritx-7}} \\ \cline{2-5} 
                                 & \textit{p-max}         & \textit{h-min}        & \textit{p-max}      & \textit{h-min}        \\ \hline
MCV                              & 0.501841      & 0.994698     & 0.501400       & 0.995966     \\
Collision                        & 0.527344      & 0.923184     & 0.521484     & 0.939304     \\
Markov                           & 4.28E-39      & 0.995748     & 3.64E-39     & 0.997594     \\
Compression                      & 0.5           & 1            & 0.5          & 1            \\
t-Tuple                          & 0.519390       & 0.945111     & 0.529343     & 0.917726     \\
LRS                              & 0.519355      & 0.945206     & 0.502963     & 0.991475     \\
Multi-MCW                        & 0.501042      & 0.998657     & 0.501141     & 0.996713     \\
Lag                              & 0.500465      & 0.998567     & 0.501683     & 0.995153     \\
Multi-MMC                        & 0.500630       & 0.998183     & 0.500566     & 0.998368     \\
LZ78Y                            & 0.501705      & 0.99509      & 0.501028     & 0.997038     \\ \hline
\multicolumn{5}{l}{* The p-value is the average of the p-values of all subtests.}
\end{tabular}
}
\vspace{-5 pt}
\end{table}
\subsubsection{AIS-31 Test}
AIS-31 is another test suit to evaluate the quality of random numbers. This test consists of nine sub-test items from T0 to T8. We collect 7,200,000 bits of random numbers for each FPGA. As shown in Table~\ref{Table 5}, the test results of AIS-31 exhibit that the random sequences generated by our DH-TRNG pass all test items.%\cite{AIS-31}
\begin{table}[htbp]
\centering
% \vspace{-3 pt}
\caption{Results of AIS-31 Test.}
\label{Table 5}
\vspace{-7 pt}
\tabcolsep=2.7pt
\scalebox{0.9}
{
\begin{tabular}{ccc}
\hline
AIS-31                & \textbf{Virtex-6} & \textbf{Artix-7} \\ \hline
Disjointness Test (T0)         & Pass     & Pass    \\
Monobit Tests (T1)*             & 100\%     & 100\%    \\
Poker Tests (T2)*               & 100\%     & 100\%    \\
Run Tests (T3)*                 & 100\%     & 100\%    \\
Long Run Test (T4)*             & 100\%     & 100\%    \\
Autocorrelation Test (T5)*      & 100\%     & 100\%    \\
Uniform Distribution Test (T6) & Pass     & Pass    \\
Multinomial Distributions (T7) & Pass     & Pass    \\
Entropy Test (T8)              & Pass     & Pass    \\ \hline
\multicolumn{3}{l}{* The pass rate of this test item.}
\end{tabular}
}
\vspace{-5 pt}
\end{table}

\subsection{Restart Test}
In the restart test, we enable the proposed DH-TRNG and sample the first 32-bit random data six times at the same condition. The obtained data are 0X8E8F7BE6, 0XD448223A, 0X2ED82918, 0X79DA4E4B, 0X51A602A9, and 0XDB9E49EC, respectively. The results show that all sets of sequences are different, which proves that our TRNG is unrepeatable and has true randomness characteristics.

\subsection{Deviation Test}
\begin{figure}[htbp]
     \centering
     \includegraphics[scale=0.64]{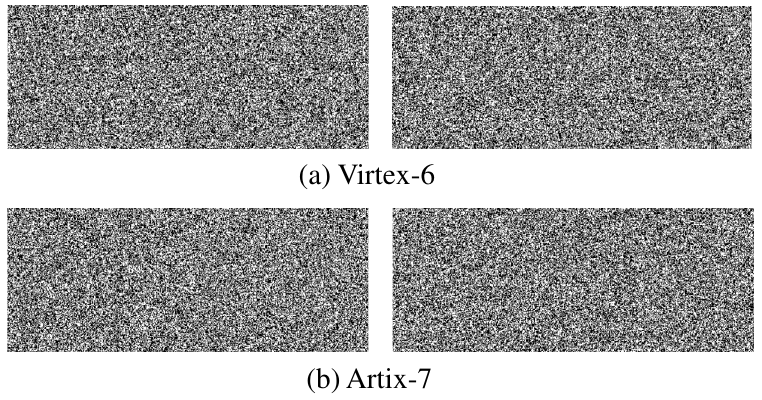}
     \vspace{-7 pt}
     \caption{Bitstream image.}
     \label{fig 7}
  \vspace{-10 pt}
 \end{figure}

In ideal conditions, a TRNG should have an equal probability of generating 0 and 1. The bias degree of a sequence not only impacts its randomness but also renders a TRNG vulnerable. We quantify the bias degree of random numbers using the following equation:
% \vspace{-7 pt}
\begin{equation}
Bias=\frac{\left | N_{1}-N_{2}   \right | }{ N_{1}+N_{2}  } \times 100 \%,
\end{equation}
where $N_{1}$ and $N_{2}$ denote the counts of 1 and 0, respectively. To evaluate our proposed TRNG, we conducted a statistical analysis using 10 sets of generated 1 Mbit random data. The statistical deviations for each FPGA are 0.0075\% (Virtex-6) and 0.0069\% (Artix-7), respectively. 
Moreover, we graph the random sequences into a bitstream image, as shown in Figure \ref{fig 7}, in which `1' is denoted by a black pixel in the left image and a white pixel in the right image. The uniform distribution of black and white pixels proves that the random numbers generated by our proposed TRNG are uniform and unbiased.
 % \vspace{-17 pt}
 \vspace{-8 pt}
\subsection{Autocorrelation Test}

\begin{figure}[htbp]
    \centering
    \includegraphics[scale=0.51]{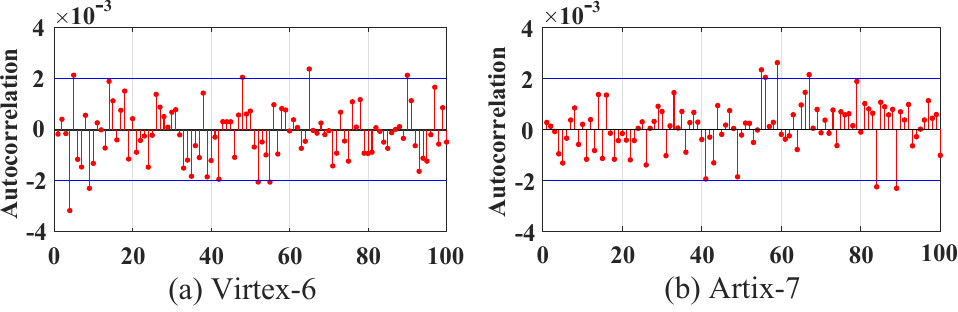}
    \vspace{-4 pt}
    \caption{Results of autocorrelation function test.}
    \label{fig 8}
\vspace{-4 pt}
\end{figure}

The autocorrelation test uses the autocorrelation function (ACF) to test whether a TRNG can produce independent random numbers. According to Karl Pearson's statistical standard, a correlation coefficient below 0.3 indicates that random sequences are uncorrelated. We set the lag coefficient from 1 to 100 and calculate the autocorrelation of 1Mbit random sequences. As shown in Figure~\ref{fig 8}, the test results indicate a low autocorrelation, demonstrating our DH-TRNG is resilient to correlation analysis attacks.
% \vspace{-5 pt}
\vspace{-3 pt}
\subsection{Process, Voltages, and Temperatures (PVT) Test}
\begin{figure}[htbp]
    \centering
    \includegraphics[scale=0.40]{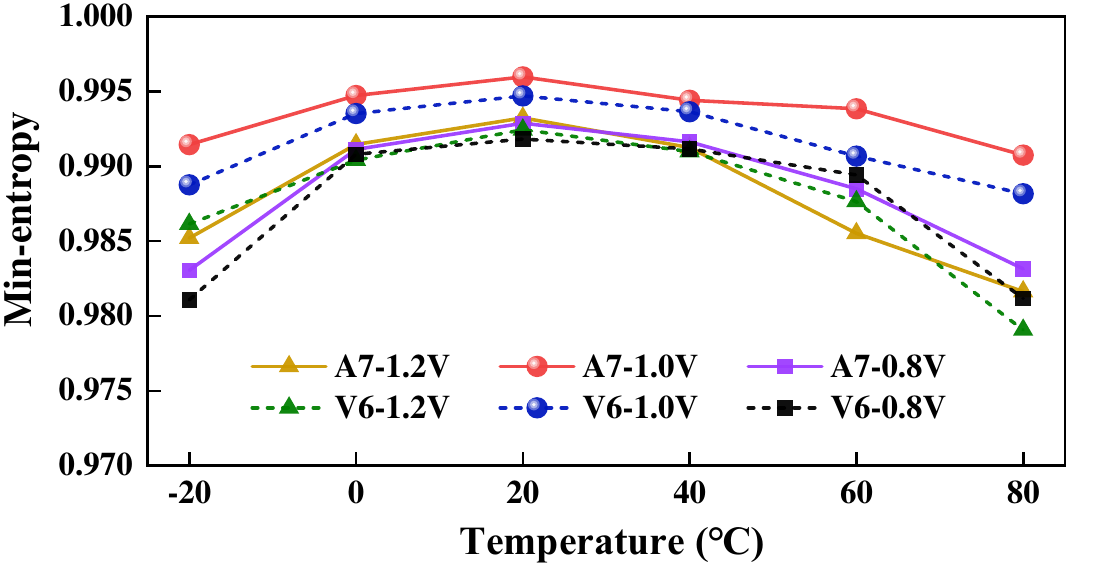}
    \vspace{-3 pt}
    \caption{Results of PVT test.}
    \label{fig 9}
    \vspace{-5 pt}
\end{figure}

The proposed DH-TRNG is evaluated across a range of temperatures (-20℃ $\sim$ 80℃) and voltages (0.8V $\sim$ 1.2V) on different process FPGAs. We collect 100 different sequences with 1Mbit in each set of experiments for the NIST SP800-90B test and calculate the min-entropy. 
As shown in Figure~\ref{fig 9}, the proposed TRNG has the largest min-entropy at 20℃ and 1.0V.
Furthermore, when the temperature and voltage vary, the min-entropy presents a slight decrease but remains consistently high, proving our DH-TRNG also has sufficient robustness to PVT fluctuations.
% \vspace{-6 pt}
\vspace{-5 pt}
\subsection{Comparison with Prior Arts}
For a fair and comprehensive evaluation, we compare the proposed design with the state-of-the-art TRNGs in terms of throughput, area, and power on Xilinx Artix-7 FPGA. As shown in Table~\ref{Table 6}, references  \cite{TCAS2_2022, FPL2020, TCAS1_2022trot, TC2023} have low power consumption but suffer from poor throughput. References  \cite{TCAS1_2021, TCAS2_2022Frustaci, DAC2023} have enhanced throughput, and the $\mathrm{\frac{Throughput}{Slices\cdot Power}}$ reaching 432.97 in \cite{DAC2023}. 
However, they generally need a high area, and the throughput needs further improvement. Different from them, our proposed entropy units and presented reinforcement strategies sufficiently exploit randomness with minimal hardware area to meet high-frequency sampling. 
It causes our DH-TRNG to stand out with the highest throughput of 620Mbps while incurring only 8 slices and a total power of 0.068W. Compared with the prior arts, DH-TRNG exhibits absolute dominance in throughput and has the highest $\mathrm{\frac{Throughput}{Slices\cdot Power} }$ of 1139.7. Thus, it achieves both ultra-high throughput and area-energy efficiency.
\vspace{-2 pt}
\begin{table}[htbp]
\centering
\renewcommand{\arraystretch}{1} % 设置行距为1.5倍，默认为1.0
\caption{Comparison in terms of throughput, area, and power.}
\label{Table 6}
\vspace{-7 pt}
\tabcolsep=1.7pt
\scalebox{0.9}
{
\begin{tabular}{ccccccc}
\hline
Design             & LUTs & DFFs & Slices & \begin{tabular}[c]{@{}c@{}}Throughput\\ (Mbps)\end{tabular} & \begin{tabular}[c]{@{}c@{}}Power*\\ (W)\end{tabular} & $\mathrm{\frac{Throughput}{Slices \cdot Power} }$       \\ \hline
FPL'20 \cite{FPL2020}     & 40   & 29   & 10     & 1.91                                                        & 0.043                                               & 4.44            \\
TCASII'21 \cite{TCAS2_2022}     & 4    & 3    & 1      & 0.76                                                        & 0.025                                               & 30.40            \\
TCASI'21 \cite{TCAS1_2021}     & 56   & 19   & 18     & 100                                                         & 0.068                                               & 81.70            \\
TCASI'22 \cite{TCAS1_2022trot} & 32   & 55   & 33     & 12.5                                                        & 0.063                                               & 6.01            \\
TCASII'22 \cite{TCAS2_2022Frustaci}  & 38   & 121  & 38     & 300                                                         & 0.119                                               & 66.34           \\
TC'23 \cite{TC2023}        & 152  & 16   & 40     & 1.25                                                        & 0.023                                               & 1.36            \\
DAC'23 \cite{DAC2023}       & 24   & 33   & 13     & 275.8                                                       & 0.049                                               & 432.97          \\
\textbf{This work} & 23   & 14   & 8      & \textbf{620}                                                & 0.068                                               & \textbf{1139.7} \\ \hline
\multicolumn{7}{l}{* The power of all architectures are conducted on Xilinx Artix-7.}
\vspace{-7 pt}
\end{tabular}}
\vspace{-4 pt}
\end{table}

\vspace{-12 pt}
\section{Conclusion}
\label{section: Conclusion}
This paper presents a dynamic hybrid TRNG with ultra-high throughput and area-energy efficiency. The proposed dynamic hybrid entropy unit sufficiently exploits both jitters and metastability to enhance randomness for high-frequency sampling. The entropy is further bolstered by the introduced coupling and feedback strategies, enabling the DH-TRNG to pass both NIST and AIS-31 tests incurring only 8 slices without any post-processing. Moreover, our design is portable across different process FPGAs and supports fully automated placement and routing. Compared with state-of-the-art TRNGs, the proposed DH-TRNG achieves the highest throughput and $\mathrm{\frac{Throughput}{Slices\cdot Power} }$. 

%future work

\vspace{-1 pt}
% \section{Acknowledgments}
\begin{acks}
This work is supported by the National Natural Science Foundation of China (No. U20A20202 and  62122023) and the Hunan Provincial Key Research and Development Project (No. 2023GK2011).
\end{acks}

%% The next two lines define the bibliography style to be used, and
%% the bibliography file.
%\bibliographystyle{unsrt}
%\bibliographystyle{ACM-Reference-Format}
%\bibliography{sample-base}

\vspace{-1 pt}
\begin{thebibliography}{00}
%\scriptsize
\bibitem{TIFS2018} N. D. Truong et al., “Machine learning cryptanalysis of a quantum random number generator,” \textit{IEEE TIFS}, pp. 403–414, 2018.

\bibitem{JSSC2022} R. Zhang et al., "A 0.186-pJ per bit latch-based true random number generator featuring mismatch compensation and random noise enhancement." \textit{IEEE JSSC}, pp. 2498–2508, 2022.

%\bibitem{DAC2015} V. Rozic et al., ``Highly Efficient Entropy Extraction for True Random Number Generators on FPGAs," in \textit{ACM/IEEE DAC}, 2015.

\bibitem{DAC2023} Z. Lu et al., “An FPGA-Compatible TRNG with Ultra-High Throughput and Energy Efficiency,” in \textit{DAC}, 2023.

\bibitem{Cui} J. Cui et al., “Design of true random number generator based on multi-stage feedback ring oscillator,” \textit{IEEE TCASII}, pp. 1752-1756, 2021.

\bibitem{Jitter} A. Hajimiri et al., “Jitter and phase noise in ring oscillators,” \textit{IEEE JSSC}, 1999.

\bibitem{JSSC2008} C. Tokunaga et al., “True random number generator with a metastability-based quality control,” \textit{IEEE JSSC}, pp. 78-85, 2008.

\bibitem{CHES2011} M. Majzoobi et al., “FPGA-based true random number generation using circuit metastability with adaptive feedback control,” in \textit{CHES}, 2011.

\bibitem{TODAES2023} Q. Peng et al., “A Compact TRNG design for FPGA based on the Metastability of RO-Driven Shift Registers,” \textit{ACM TODAES}, 2023.

\bibitem{TCAS1_2022} R. Della Sala et al., “High-throughput FPGA-compatible TRNG architecture exploiting multistimuli metastable cells,” \textit{IEEE TCASI}, 2022.

\bibitem{XOR2008} K. Wold et al., “Analysis and enhancement of random number generator in fpga based on oscillator rings,” in \textit{ IEEE International Conference on Reconfigurable Computing and FPGAs}, 2008.

\bibitem{TCAS1_2017} Y. Liu et al., "A bias-bounded digital true random number generator architecture." \textit{IEEE TCASI}, pp. 133-144, 2017.

%\bibitem{TC2006} J. D. Golic., "New methods for digital generation and postprocessing of random data." \textit{IEEE TC}, 2006.

%\bibitem{NIST22} A. Rukhin et al., “A statistical test suite for random and pseudorandom number generators for cryptographic applications,” Booz-allen and hamilton inc mclean va, Tech. Rep., 2001.

%\bibitem{NIST90B} M. S. Turan et al., “Recommendation for the entropy sources used for random bit generation,” \textit{NIST Special Publication}, 2018.

%\bibitem{AIS-31} W. Killmann et al., “A proposal for: Functionality classes for random number generators,”  \textit{ser. BDI, Bonn}, 2011, pp. 1–133.

\bibitem{FPL2020} N. Fujieda, “On the feasibility of tero-based true random number generator on xilinx fpgas,” in \textit{FPL}, 2020.

\bibitem{TCAS2_2022} R. Della Sala et al., “A novel ultra-compact FPGA-compatible TRNG architecture exploiting latched ring oscillators,” \textit{IEEE TCASII}, 2022.

\bibitem{TCAS1_2021} X. Wang et al., “High-throughput portable true random number generator based on jitter-latch structure,” \textit{IEEE TCASI}, 2021.

\bibitem{TCAS1_2022trot} M. Gruji´c et al., “Trot: A three-edge ring oscillator based true random number generator with time-to-digital conversion,” \textit{IEEE TCASI}, 2022.

\bibitem{TCAS2_2022Frustaci} F. Frustaci et al.,  “A high-speed fpga-based true random number generator using metastability with clock managers,” \textit{IEEE TCASII}, 2022.

\bibitem{TC2023} K. Pratihar et al., “Birds of the Same Feather Flock Together: A Dual-Mode Circuit Candidate for Strong PUF-TRNG Functionalities.” \textit{IEEE TC}, 2023.























%\bibitem{TC2019} P. Poudel, B. Ray, and A. Milenkovic, “Microcontroller TRNGs using perturbed states of NOR flash memory cells,” \textit{IEEE Trans. Comput.}, vol. 68, no. 2, pp. 307–313, Feb. 2019.

%\bibitem{hilbert2011world} M. Hilbert, P. López, “The world’s technological capacity to store, communicate, and compute information,” \textit{Science}, vol. 332, no. 6025, pp. 60–65, 2011.

%\bibitem{qiu2020practical} S. Qiu, D. Wang, G. Xu, and S. Kumari, “Practical and provably secure three-factor authentication protocol based on extended chaotic-maps for mobile lightweight devices,” \textit{IEEE Transactions on Dependable and Secure Computing}, 2020.











%\bibitem{PLLTRNG} B. Colombier et al., ``Backtracking Search for Optimal Parameters of a PLL-based True Random Number Generator," in \textit{IEEE DATE}, 2020.

%\bibitem{GarbledCircuits} A. Ben-Efraim et al., ``Large Scale, Actively Secure Computation from LPN and Free-XOR Garbled Circuits," in \textit{EUROCRYPT}, 2021.

%\bibitem{ROTRNG2010} S.-K. Yoo et al., ``Improving the Robustness of Ring Oscillator TRNGs," \textit{ACM Trans. Reconfigurable Technol. Syst.}, 2010.

%\bibitem{NISTIR} G. Alagic et al., ``Status Report on the Third Round of the NIST Post-Quantum Cryptography Standardization Process," \textit{NIST Interagency/Internal Report}, 2022.

%\bibitem{CHES2013} A. Cherkaoui et al., ``A very high speed true random number generator with entropy assessment," in \textit{IACR CHES}, 2013.

%\bibitem{DATE2012} A. Cherkaoui et al., ``Comparison of Self-Timed Ring and Inverter Ring Oscillators as Entropy Sources in FPGAs," in \textit{IEEE DATE}, 2012.

%\bibitem{TII2016} H. Martin et al., ``A new TRNG based on coherent sampling with self- timed rings," \textit{IEEE Trans. Ind. Informat.}, 2016.

%\bibitem{TCASI2017} Y. Liu et al., ``A bias-bounded digital true random number generator architecture," \textit{IEEE TCAS-I}, 2017.

%\bibitem{TCASI2021} X. Wang et al., ``High-Throughput Portable True Random Number Generator Based on Jitter-Latch Structure," \textit{IEEE TCAS-I}, 2021.


\vspace{5 pt}
\end{thebibliography}

%%
%% If your work has an appendix, this is the place to put it.
\appendix

\end{document}